\documentclass[fleqn,twoside]{article}
\usepackage{espcrc2}

\usepackage{graphicx}
\usepackage[figuresright]{rotating}

\newcommand{\Antares}{\textsc{Antares }}

\newcommand{\AmS}{{\protect\the\textfont2
  A\kern-.1667em\lower.5ex\hbox{M}\kern-.125emS}}

\title{\Antares completed: First selected results}

\author{E. Presani\address[Nikhef]{Nikhef, 
        Science Park, \\ 
        Kruislaan 409, 1098 SJ Amsterdam, The Netherlands}%
	\thanks{on behalf of the \Antares collaboration}
        }
       
\begin{document}

\begin{abstract}
In May 2008, the \Antares collaboration has completed the construction of the first
 deep sea neutrino telescope in the Northern hemisphere.
\Antares is a 3D array of 900 photomultipliers held in the sea by twelve mooring
 lines anchored at a depth of 2500 m in the Mediterranean Sea 40 km off the southern French coast.
The detection principle is based on the observation of \u{C}erenkov light
 induced by charged particles produced in neutrino interactions in the
 matter surrounding the detector.

\vspace{1pc}
\end{abstract}

\maketitle

\section{Introduction}

Detecting cosmic neutrinos will allow us to obtain information from the inner core of distant sources. Because of their weak coupling to matter, they are not affected neither by massive obstacles nor by background radiation in their way and, as they are not charged, they are not affected by magnetic fields either, and consequently point back to their source. Candidates for neutrino sources are e.g. active galactic nuclei (AGNs) and transient sources like gamma ray bursts (GRB). Hadrons accelerated in these sources interact with the ambient matter and dense photon fields and create pions, which produce neutrinos in their decay.
\begin{figure}[!h]
\includegraphics[width = 17 pc]{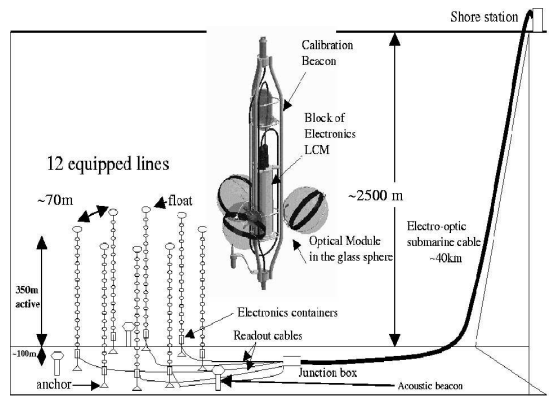} \label{layout}
\caption{\Antares detector. A storey is shown in the enlarged view.}
\end{figure}
The detection principle of a neutrino telescope such as \Antares relies on the observation of \u{C}erenkov light produced by neutrino induced muons by a three dimensional array of photodetectors. The higher the neutrino energy is, the smaller is the angle between the direction of the neutrino and the produced muon.

Cosmic rays interacting with the nuclei in the atmosphere cause a cascade of secondary particles, amongst which high energy muons that constitute an intense source of background in the detector. To suppress this background ($\sim 1$ per second at the reconstruction level), \Antares is optimized to detect upward going muons produced by neutrinos within or in the vicinity of the detector after traveling through the Earth. Other sources of background are upward going atmospheric neutrinos ($\sim 10/$ day), which will have the same signature in the \Antares detector as the expected cosmic signal. 
Muons generated by $\nu_\mu$ interactions produce a long track in the detector since their interaction length is considerably long in water.

\section{The ANTARES detector}
The \Antares neutrino telescope, deployed at a depth of 2500 m in the Mediterranean Sea, 40~km off the coast of
La Seyne-sur-Mer, France, is composed of twelve mooring strings, holding a total of 900 photomultipliers. \\
From January to December 2007 five lines were operational, during that month the detector was increased to ten operational lines and in May 2008 the twelve-line detector was completed. A schematic description of the detector can be found in figure \ref{layout}.                                                                               Horizontally, lines are separated by 60-75 m. Each line holds 25 storeys with a triplet of Optical Modules. Each optical module consists of a pressure resistant-glass sphere housing a photomultiplier looking downward at 45 degrees from the horizontal.

The performance of the detector depends on the properties of light propagation 
at the \Antares site, and are summarized in Ref.\cite{ANT1}. The absorption length in the blue (470 nm) is about 60~m 
while the effective scattering length is about 250~m. 
The overall time calibration is better than 0.5~ns ensuring the possibility to achieve
an angular resolution at the level of $0.3^\circ$ for muons above 10 TeV \cite{ANT3}.
\Antares is equipped with a positioning system which measures the line displacements with a precision
on the relative location better than 10 cm.
\begin{figure}[!h]
\includegraphics[width=18 pc]{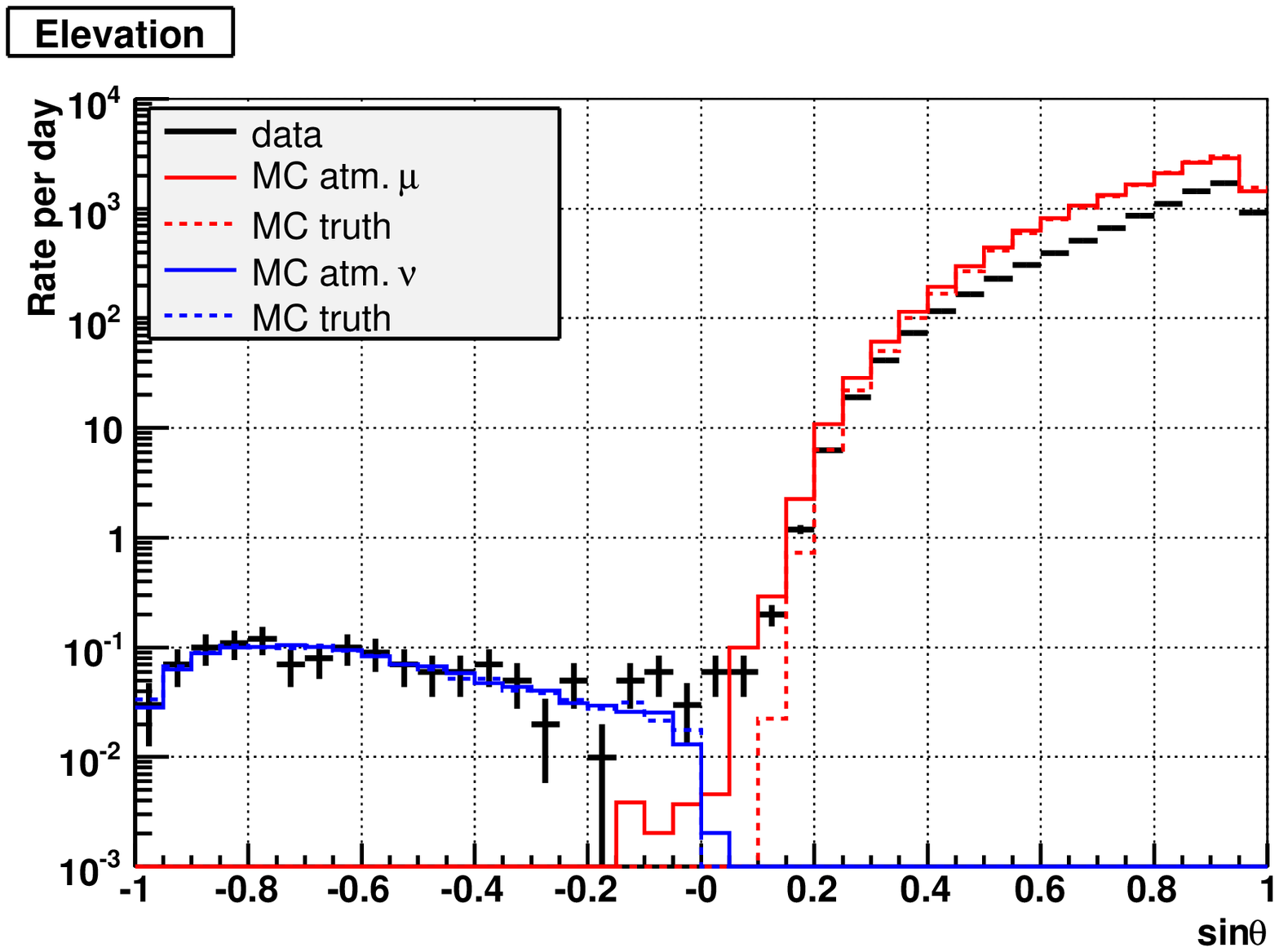} \\
\includegraphics[width=18 pc]{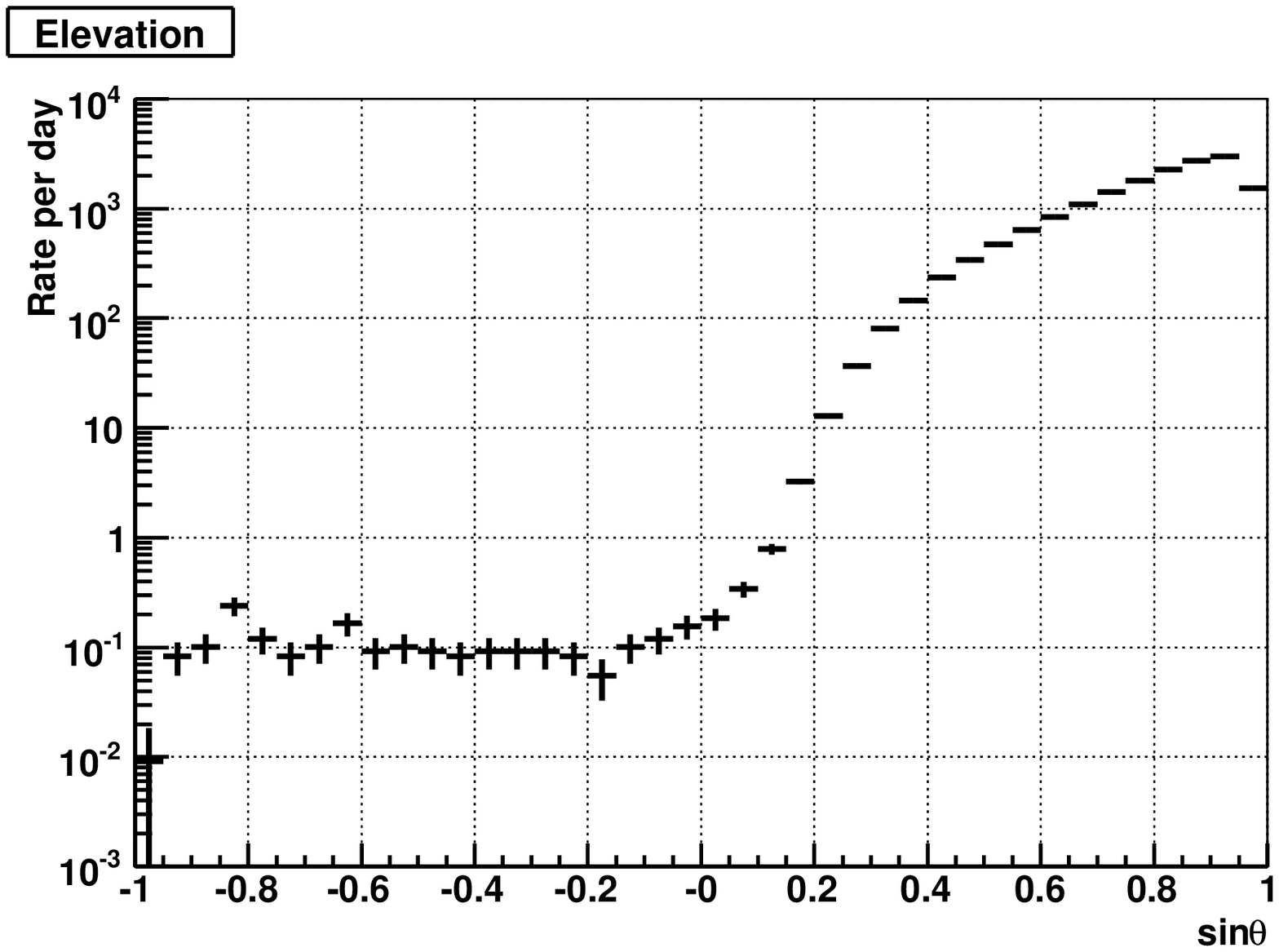}
\caption{Above: Elevation angle distribution for 120.5 days of 5-line data, compared to MonteCarlo.
Below: Same distribution for 10-line and 109 days of data taking}
\end{figure}\label{zenith}
The zero level trigger condition (L0) accepts only hits above a threshold of 1/3 of a photoelectron (pe) to be sent to shore. 
The first level trigger condition (L1) is formed by a local coincidence of 2 L0 in a storey within 
a 20 ns window or a large pulse above a few pe, usually 3 pe. 
To further reduce the trigger rate a causality relation is required between at least 5 L1 hits
compatible with a light signal produced by an ultra-relativistic muon. The measured
muon trigger rate is about 3 Hz in the 10-line detector. In addition to this standard trigger, other special triggers
have been developed:for example a gamma-ray bursts trigger and a directional trigger pointing at the Galactic Centre. 
The filter is necessary since sea water is an optical noisy environment due to bioluminescence
and $\beta$ decay of $^{40}$K producing electrons. In the 10-line configuration we observed that
the optical background rate is about 60 kHz during 80\% of the time. 
In addition to the optical background, the limited density of the photosensors in \Antares
can lead to misreconstructed events that can mimic upward going neutrino 
induced muons. This is particularly the case for events close to the border or for large muon
bundles where the time pattern might be confusing. 

\section{First Results}
In Fig \ref{zenith} the elevation angle distribution for the 5 and 10-line detector is shown; the top figure also shows the comparison with Monte Carlo simulation. The tracks of the muons shown in the figure are reconstructed using a linear fit algorithm that selects hits on a time basis. 

For the search of steady point sources new unbinned methods have been developed based on event pattern recognition.They appear to be up to 50\% more powerful than the more traditional binned method based on a predefined angular window \cite{ANT2}.
Time signatures, such as in the case of GRBs, can also help to discriminate atmospheric neutrinos and
residual atmospheric muon backgrounds.
\begin{figure}[!h]\label{point}
\includegraphics[width=20 pc, height = 17pc]{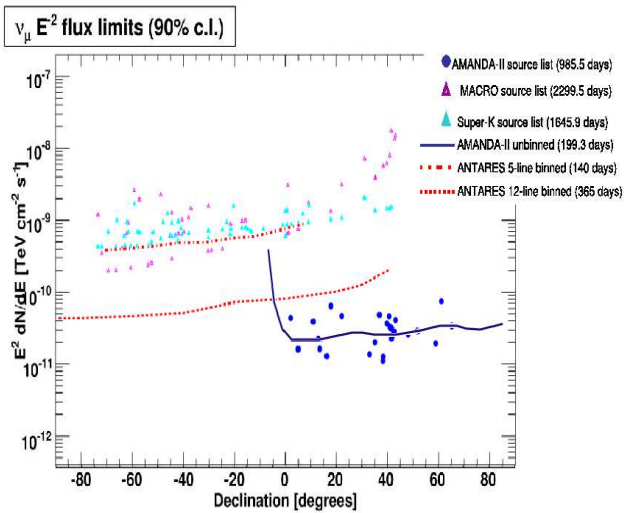}
\caption{Sensitivity of 5-line of \Antares (binned method) after 140 days to $E^{-2}$ flux of neutrinos vs declination.}
\end{figure}
    The 5-line data have been used to estimate the discovery potential of \Antares for point like sources. The sensitivity for the all-sky search is shown
Fig. \ref{point}. Although this is not the full detector configuration and the algorithm for muon track reconstruction is expected to be further improved, the sensitivity for 140 days with binned method is already at the level of many years
of detection of the first generation experiments (see \cite{SKAM} \cite{MACRO}), with unbinned methods the sensitivity is expected to improve. The estimated result for a full year of data taking with the 12 lines detector is also shown.

\section{Summary}
 Since May 2008 the full twelve line \Antares detector is completed and taking data. First results with the 5 and 10-line configurations are promising as they are comparable with previous experiments, and predictions are optimistic for improved results. The position of \Antares in the Mediterranean Sea will provide complementary data to those of the Amanda and IceCube detectors, situated in the South Pole.
 
Different triggers and reconstruction methods are being developed and will be available for the analysis of the twelve-line data.

\end{document}